\documentclass[12pt, a4paper]{article}
\usepackage{cite}

\usepackage[utf8]{inputenc}
\usepackage[english]{babel}
\usepackage[margin=1in]{geometry}
\usepackage{graphicx}
\usepackage{amsmath, amsfonts, amssymb}
\usepackage{hyperref}
\usepackage{tikz} 
\usetikzlibrary{decorations.pathmorphing}
\usepackage{authblk} 
\usepackage{subcaption} 
\usepackage{float} 
\title{Quantum Thermometry of External Phonon Reservoirs in Driven Open Quantum Systems}

\author{Yildiz Berk Ates}
\affil{Department of Physics, University of Warwick, Coventry CV4 7AL, United Kingdom \\
Email: \href{mailto:yildiz.berk@hotmail.com}{yildiz.berk@hotmail.com}, \href{mailto:yildiz.ates-berk@warwick.ac.uk}{yildiz.ates-berk@warwick.ac.uk}}

\date{April 2026} 

\begin{document}
\maketitle

\begin{abstract}
We investigate the non-monotonic temperature sensitivity of a coherently driven two-level quantum system coupled to an Ohmic phonon environment. By employing a unitary polaron transformation, we account for phonon-induced renormalization effects that go beyond the standard weak-coupling approximations. Our analysis reveals that the Quantum Fisher Information (QFI) exhibits a prominent peak at an intermediate system-environment coupling strength, identifying an optimal regime for thermal sensing. This behavior emerges from a fundamental competition between environment-induced dissipation enhancement and the exponential suppression of system parameters due to phonon dressing. We demonstrate that while thermometric precision vanishes in both the ultra-weak and strong coupling limits, a properly tuned nonequilibrium steady state can significantly enhance sensitivity. These results suggest that environmental interactions, often viewed as detrimental decoherence sources, can be engineered as a resource to optimize the performance of solid-state quantum thermometers.

\end{abstract}

\section{Introduction}

The quest for high-precision thermometry at the nanoscale has become a cornerstone of modern quantum technologies, driven by the rapid advancement of solid-state platforms and quantum sensing \cite{mehboudi2019, correa2015, depasquale2018}.Central to this field is the Quantum Fisher Information (QFI), which establishes the fundamental precision limit for temperature estimation via the quantum Cramér-Rao inequality \cite{braunstein1994, paris2009}. While quantum thermometry has traditionally relied on probes reaching thermal equilibrium with their environment, recent efforts have shifted toward leveraging nonequilibrium steady states to enhance sensitivity \cite{breuer2002, weiss2012, schaller2014}.
However, a significant portion of the literature remains confined to the weak-coupling regime, where the environment is treated as a minor perturbation. In many physical systems, such as quantum dots or NV centers in diamond, the interaction with the surrounding phonon bath can be substantial, leading to complex dressing effects that cannot be captured by standard master equations \cite{leggett1987, nazir2018, silbey1984}.These interactions do not merely cause decoherence; they fundamentally renormalize the system's energy scales and dissipation rates.

In this work, we move beyond the weak-coupling approximation by employing a polaron transformation to investigate the temperature sensitivity of a coherently driven two-level system. Our approach allows us to explore the intermediate and strong-coupling regimes where the interplay between coherent driving and an Ohmic phonon environment dictates the probe's steady-state response. We show that the QFI is not a monotonic function of the coupling strength; instead, it reaches a maximum at an optimal coupling point. This peak emerges from a delicate balance between environment-induced dissipation and the exponential suppression of the effective drive due to phonon dressing. By identifying this optimal regime, we provide a blueprint for engineering more robust quantum thermometers in structured environments.

\newpage
\section{Model}
\begin{figure}[ht]
    \centering
    \begin{tikzpicture}[scale=1.2]
        \draw[thick] (-0.5,0) -- (0.5,0) node[right] {$|g\rangle$};
        \draw[thick] (-0.5,1.5) -- (0.5,1.5) node[right] {$|e\rangle$};
        
        \draw[<->, >=stealth, blue, thick] (0,0.1) -- (0,1.4) node[midway, left] {$\omega_0$};
        
        \draw[<->, >=stealth, red, thick, bend left=45] (-0.6,0.2) to node[left] {$\Omega$} (-0.6,1.3);
        
        \draw[gray, thick, dashed] (2,0.75) ellipse (1.2cm and 0.8cm);
        \node at (2,0.75) {Phonon Bath};
         \node[below] at (2,0.3) {$J(\omega), T$};

        \draw[decorate, decoration={snake, amplitude=1mm, segment length=2mm}, thick] (0.6,0.75) -- (1.5,0.75) node[midway, above] {$\eta$};
        
    \end{tikzpicture}
  \caption{Schematic representation of the physical model: A two-level quantum system (qubit) with transition frequency $\omega_0$ is coherently driven by an external field with amplitude $\Omega$. The system is coupled to an Ohmic phonon environment, characterized by spectral density $J(\omega)$ and temperature $T$, with an interaction strength $\eta$.}

    \label{fig:system_schema}
\end{figure}
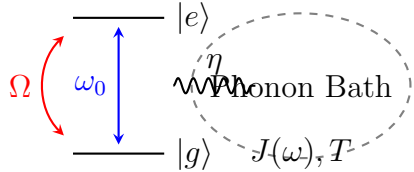

We consider a coherently driven two-level quantum system. The system Hamiltonian is given by
\begin{equation}
H_S = \frac{\omega_0}{2}\sigma_z + \Omega \sigma_x
\end{equation}
where $\omega_0$ denotes the bare transition frequency and $\Omega$ represents the driving amplitude.

The system is coupled to a bosonic environment, modeled as a phonon bath with an Ohmic spectral density of the form
\begin{equation}
J(\omega) = \eta \omega^s e^{-\omega/\omega_c}, \qquad s=1.
\end{equation}
Here, $\eta$ denotes the system--bath coupling strength and $\omega_c$ is the cutoff frequency.

Such spectral densities are commonly used to describe dissipative environments in solid-state systems, where phonon modes provide the dominant source of decoherence.

\section{Open System Dynamics}

Under the Born–Markov approximation and assuming weak system–environment memory effects, the dynamics of the reduced density matrix can be described by a Lindblad master equation [1,2]:
\begin{equation}
\dot{\rho} = -i[H_S,\rho] + \gamma(\eta,T)\mathcal{D}[\sigma_-]\rho .
\end{equation}
The dissipator superoperator is defined as
\begin{equation}
\mathcal{D}[A]\rho = A\rho A^\dagger - \frac{1}{2}\{A^\dagger A,\rho\}.
\end{equation}
Within this framework, the system–environment interaction gives rise to temperature-dependent dissipation, which is determined by the spectral density and the Bose–Einstein distribution. In particular, the damping rate depends on both the coupling strength and the thermal occupation of the bath modes.
In the following section, we extend this standard description by incorporating phonon-induced renormalization effects.

\section{Phonon Dressing and Renormalization}
To go beyond the standard weak-coupling description, it is convenient to work in a displaced oscillator basis using a unitary polaron transformation \cite{jevtic2015, cavina2018, sekatski2022, guarnieri2019}.

 In this transformed frame, the effect of the phonon bath is not limited to dissipation. Instead, it also modifies the intrinsic system parameters through a renormalization mechanism.

This effect can be captured by a temperature-dependent rescaling factor, which originates from the thermal average of the bath displacement operators. It can be written in terms of the spectral density as

\begin{equation}
f(\eta,T) = \exp\left[ -\frac{1}{2} \int_0^\infty d\omega \frac{J(\omega)}{\omega^2} \coth\left(\frac{\omega}{2T}\right) \right].
\end{equation}

For an Ohmic spectral density with a high-frequency cutoff $\omega_c$, and evaluating the integral near the relevant system frequencies, this evaluates to:

\begin{equation}
f(\eta,T) = \exp\left[-\frac{\eta}{\omega_c}(2n(\omega_0,T)+1)\right],
\end{equation}

where $n(\omega_0,T) = \frac{1}{e^{\omega_0/T}-1}$ is the Bose-Einstein distribution.

This factor modifies both the effective transition frequency and the damping rate as
\begin{equation}
\omega_{\text{eff}}(\eta,T) = \omega_0 f(\eta,T),
\end{equation}
\begin{equation}
\gamma(\eta,T) = 2\pi J(\omega_0,\eta) \big(n(\omega_0,T)+1\big) f(\eta,T).
\end{equation}
In the present work, we take the driving amplitude $\Omega$ to be the bare (unrenormalized) value, while the polaron transformation is used solely to obtain the renormalized frequency $\omega_{\text{eff}}$ and the damping rate $\gamma$. The full renormalization of the drive, $\Omega_{\text{eff}} = \Omega f(\eta, T)$, is discussed in Appendix A. To isolate the primary competition between environment-induced dissipation enhancement and the exponential suppression of system parameters, we employ the bare driving amplitude $\Omega$, as its full renormalization yields secondary effects in the intermediate coupling regime. 

This structure leads to two competing effects induced by the coupling strength:
\begin{itemize}
    \item An increase in dissipation due to $J(\omega_0,\eta) \propto \eta$,
    \item An exponential suppression arising from the renormalization factor $f(\eta,T)$.
\end{itemize}

The competition between these two mechanisms forms the basis of the metrological behavior of the system.

\section{Steady-State Solution}

The system dynamics are governed by the optical Bloch equations for populations and coherences. In the polaron frame, and under the secular approximation, the equations of motion are expressed as:

\begin{align}
\dot{\rho}_{ee} &= -i\frac{\Omega}{2}(\rho_{eg}-\rho_{ge})-\gamma(\eta,T)\rho_{ee}, \label{eq:pop_evol} \\
\dot{\rho}_{eg} &= -i\omega_{eff}(\eta,T)\rho_{eg}-i\frac{\Omega}{2}(\rho_{ee}-\rho_{gg})-\frac{\gamma(\eta,T)}{2}\rho_{eg}. \label{eq:coh_evol}
\end{align}

By setting the time derivatives to zero ($\dot{\rho}_{ee} = \dot{\rho}_{eg} = 0$) and invoking the trace preservation condition ($\rho_{gg} + \rho_{ee} = 1$), we solve the resulting algebraic system to find the steady-state excited-state population:

\begin{equation}
P_{e}(\eta,T)=\frac{\Omega^{2}}{2\Omega^{2}+\gamma(\eta,T)^{2}+\omega_{eff}(\eta,T)^{2}}. \label{eq:steady_state_sol}
\end{equation}

This expression reveals that both the dissipation rate and the phonon-induced renormalization directly dictate the steady-state response, ultimately determining the probe's thermometric precision.

\subsection{Analytical Structure of the Temperature Dependence}

To better understand the origin of the non-monotonic behavior, we analyze the temperature derivative of the steady-state population.

Recalling that
\begin{equation}
P_e =
\frac{\Omega^2}
{D},
\qquad
D = 2\Omega^2 + \gamma^2 + \omega_{\text{eff}}^2,
\end{equation}
the derivative reads
\begin{equation}
\partial_T P_e
=
-\frac{\Omega^2}{D^2}
\left(
2\gamma\,\partial_T \gamma
+
2\omega_{\text{eff}}\,\partial_T \omega_{\text{eff}}
\right).
\end{equation}

The effective frequency is given by
\begin{equation}
\omega_{\text{eff}} = \omega_0 f(\eta,T),
\end{equation}
with the phonon dressing factor
\begin{equation}
f(\eta,T)
=
\exp\left[
-\frac{\eta}{\omega_c}(2n(\omega_0,T)+1)
\right].
\end{equation}

Thus,
\begin{equation}
\partial_T \omega_{\text{eff}}
=
\omega_{\text{eff}}
\left(
-\frac{\eta}{\omega_c}
\right)
2 \partial_T n(\omega_0,T).
\end{equation}

Using
\begin{equation}
\partial_T n(\omega_0,T)
=
\frac{\omega_0 e^{\omega_0/T}}
{T^2\left(e^{\omega_0/T}-1\right)^2},
\end{equation}
we see that the temperature response is strongly enhanced at low temperatures.

Similarly, the dissipation rate
\begin{equation}
\gamma \propto \eta (n+1) f(\eta,T)
\end{equation}
contains both a linear coupling enhancement and an exponential suppression.

Therefore, the QFI is governed by two competing mechanisms:

(i) growth of dissipation with increasing coupling,  
(ii) exponential suppression due to phonon dressing.

The competition between these effects gives rise to an optimal intermediate coupling regime, where the temperature sensitivity is maximized.

\subsection{Quantum Fisher Information for the Steady State}
The ultimate precision of estimating the temperature $T$ is bounded by the quantum Cramér--Rao bound, $\text{Var}(T) \geq 1 / (\nu F_Q(T))$, where $\nu$ is the number of independent measurements and $F_Q(T)$ is the Quantum Fisher Information. For a two-level system where the steady-state density matrix is diagonal in the energy eigenbasis, the QFI reduces to the classical Fisher Information evaluated over the populations. Thus, the QFI is given by:

The QFI is calculated by summing over the ground and excited state contributions:
\begin{equation}
F_Q(T) = \frac{(\partial_T P_e)^2}{P_e} + \frac{(\partial_T P_g)^2}{P_g}.
\end{equation}
Since probability is strictly conserved ($P_g = 1 - P_e$), we have the relation $\partial_T P_g = -\partial_T P_e$. Substituting this into the sum, the expression simplifies directly to:

\begin{equation}
F_Q(T) = (\partial_T P_e)^2 \left( \frac{1}{P_e} + \frac{1}{1-P_e} \right) = \frac{(\partial_T P_e)^2}{P_e(1-P_e)}.
\end{equation}

This formulation establishes a direct, computable link between the temperature derivative of the excited-state population, $\partial_T P_e$ (derived in Eq. 12), and the metrological performance of the probe. \textit{It is important to note that in the polaron-transformed frame, the off-diagonal elements of the density matrix (coherences) decay rapidly due to environment-induced dephasing. Since the steady state $\rho_{ss}$ remains approximately diagonal in the effective Hamiltonian basis, and the thermometric information is primarily encoded in the population balance, the contribution of residual coherences to the QFI is negligible, justifying the use of Eq. (20).}

\section{Results and Discussion}

In this section, we provide a numerical analysis of the temperature sensitivity. For broad applicability across different solid-state platforms, the quantum Fisher information $F_Q$ and related parameters are presented in dimensionless units, scaled relative to the bare transition frequency $\omega_0$.

\subsection{QFI as a Function of Coupling Strength}

\begin{figure}[htbp]
    \centering
    \begin{subfigure}[b]{0.48\textwidth}
        \includegraphics[width=\textwidth]{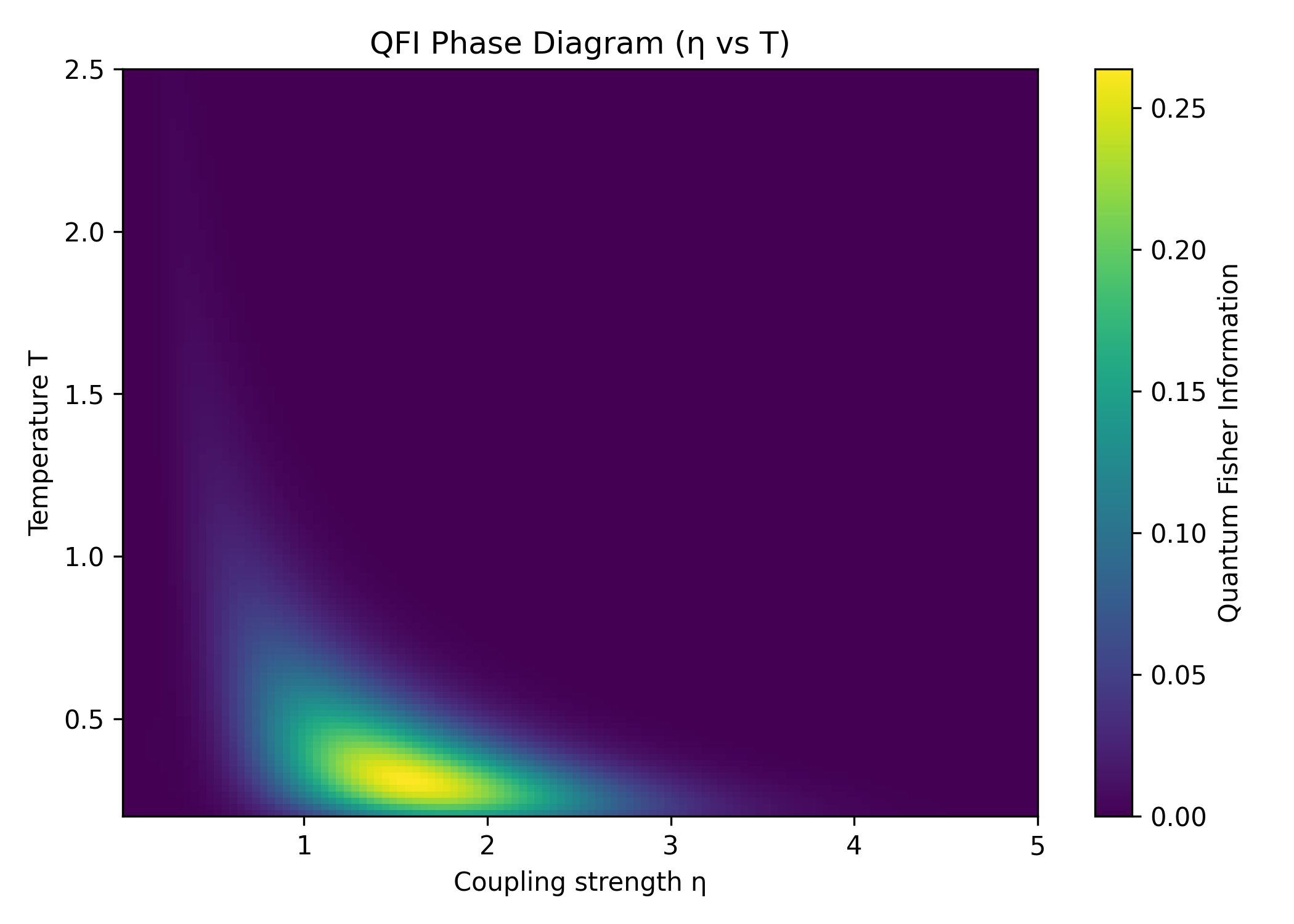}
        \caption{}
    \end{subfigure}
    \hfill
    \begin{subfigure}[b]{0.48\textwidth}
        \includegraphics[width=\textwidth]{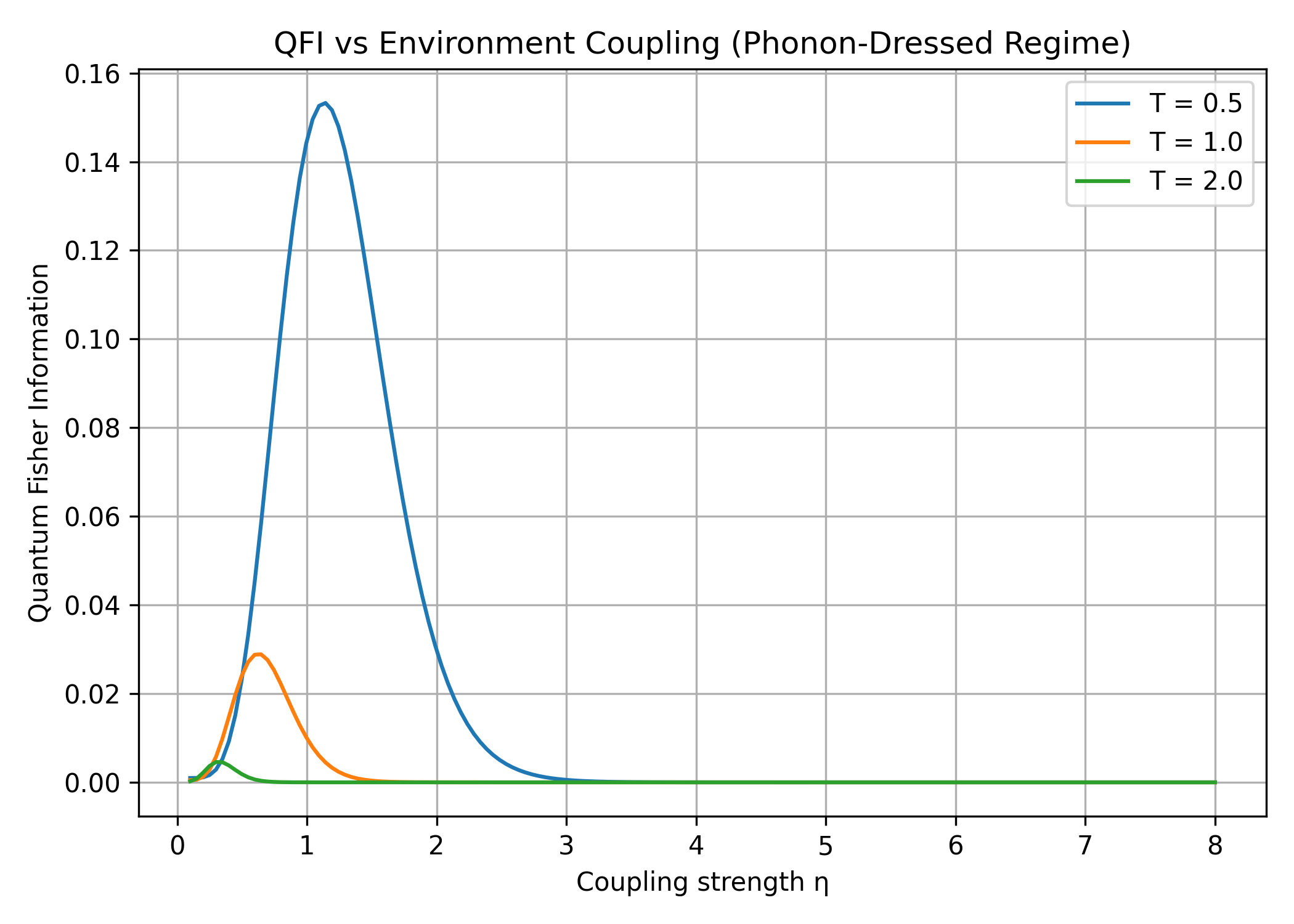}
        \caption{}
    \end{subfigure}
    \caption{{Global and local behavior of the Quantum Fisher Information (QFI).} (a) Phase diagram of the QFI as a function of temperature $T$ and coupling strength $\eta$. The prominent "island" structure highlights the optimal intermediate coupling regime where temperature sensitivity is maximized. (b) QFI as a function of $\eta$ for representative temperatures ($T = 0.5, 1.0, 2.0$). The non-monotonic behavior illustrates the competition between environment-induced dissipation and exponential phonon-induced renormalization.}
    \label{fig:qfi_behavior}
\end{figure}

The global behavior of the QFI across the parameter space of temperature $T$ and coupling strength $\eta$ is presented in the phase diagram in Figure~\ref{fig:qfi_behavior}(a). To provide a more detailed quantitative analysis, Figure~\ref{fig:qfi_behavior}(b) shows the QFI as a function of the coupling strength $\eta$ for different temperatures.

For weak coupling ($\eta \to 0$), the dissipation rate is small and the phonon dressing factor approaches unity. In this regime, the system behaves close to an isolated driven two-level system, resulting in limited temperature sensitivity. This is visible as the dark purple regions on the left side of the phase diagram.

As the coupling strength increases, dissipation becomes temperature dependent through the Bose–Einstein distribution. This enhances the derivative $\partial_T P_e$ and consequently increases the QFI. An optimal intermediate coupling regime emerges, where temperature sensitivity is maximized, forming the prominent ``island'' of high QFI values seen in Figure~\ref{fig:qfi_behavior}(a).

For strong coupling, however, the exponential phonon dressing factor significantly suppresses both the effective frequency and the dissipation rate. Consequently, the steady-state population becomes weakly dependent on temperature, and the QFI decreases. This non-monotonic behavior directly reflects the competition between linear dissipation enhancement and exponential renormalization, a transition clearly captured by the fading of the QFI island as $\eta$ increases beyond the optimal point.

This non-monotonic behavior directly reflects the competition between linear dissipation enhancement and exponential renormalization.

\subsection{QFI as a Function of Temperature}
\begin{figure}[htbp]
    \centering
    \includegraphics[width=0.8\textwidth]{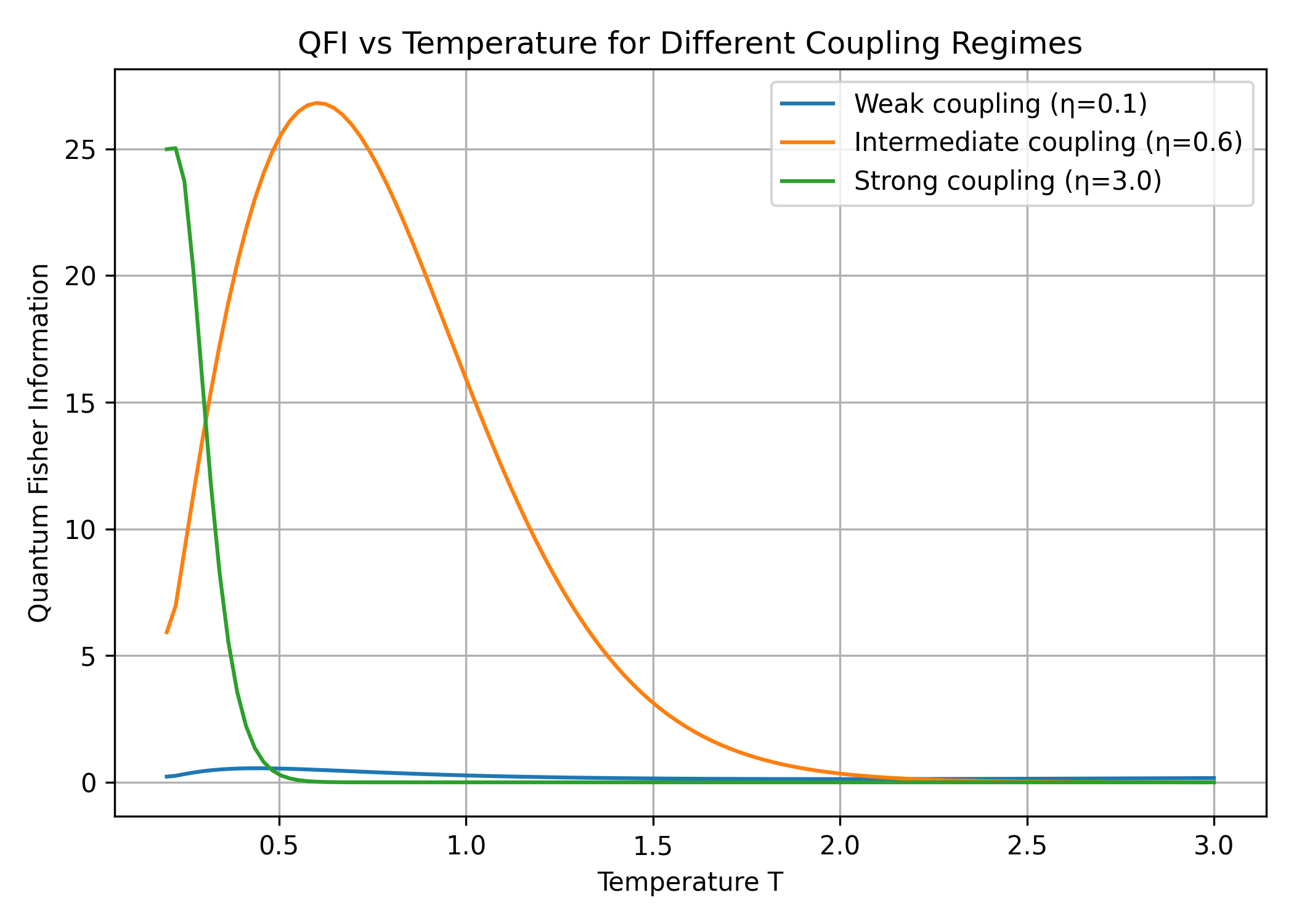}
    \caption{QFI as a function of temperature for representative weak, intermediate, and strong coupling regimes.}
    \label{fig:birinci_grafik}
\end{figure}

In Figure 3, we plot the QFI as a function of temperature for representative weak, intermediate, and strong coupling regimes.

In the weak-coupling regime, the QFI exhibits a broad and relatively small peak, indicating moderate sensitivity over a wide temperature range.

In the intermediate regime, the peak becomes sharper and higher, signaling enhanced precision around a specific optimal temperature. This regime provides the best overall metrological performance.

In the strong-coupling regime, the QFI is suppressed at high temperatures due to exponential renormalization, but remains enhanced at low temperatures where the derivative of the Bose–Einstein distribution is large.

These results demonstrate that coupling strength does not merely rescale the sensitivity, but qualitatively modifies the temperature response of the probe.As illustrated in Figure 3, the temperature dependence of the QFI varies qualitatively across different coupling regimes. In the weak-coupling regime ($\eta=0.1$),the QFI remains suppressed across the entire temperature range, as the probe is insufficiently affected by the thermal environment to provide high-precision estimation.

In the intermediate-coupling regime ($\eta=0.6$), we observe a significant enhancement in sensitivity, characterized by a prominent peak at $T \approx 0.5$. This confirms that an optimal interaction strength exists where the probe's response to temperature changes is maximized.
 In contrast, for the strong-coupling regime ($\eta=3.0$), the QFI is strongly peaked at very low temperatures but decays rapidly as temperature increases. This is due to the exponential phonon-induced renormalization f($\eta, T$) becoming dominant, which effectively "freezes" the system’s dynamics and degrades its thermometric capability at higher thermal energies.

\subsection{QFI as a Function of Cutoff Frequency}
\begin{figure}[H]
\centering
\includegraphics[width=0.8\linewidth]{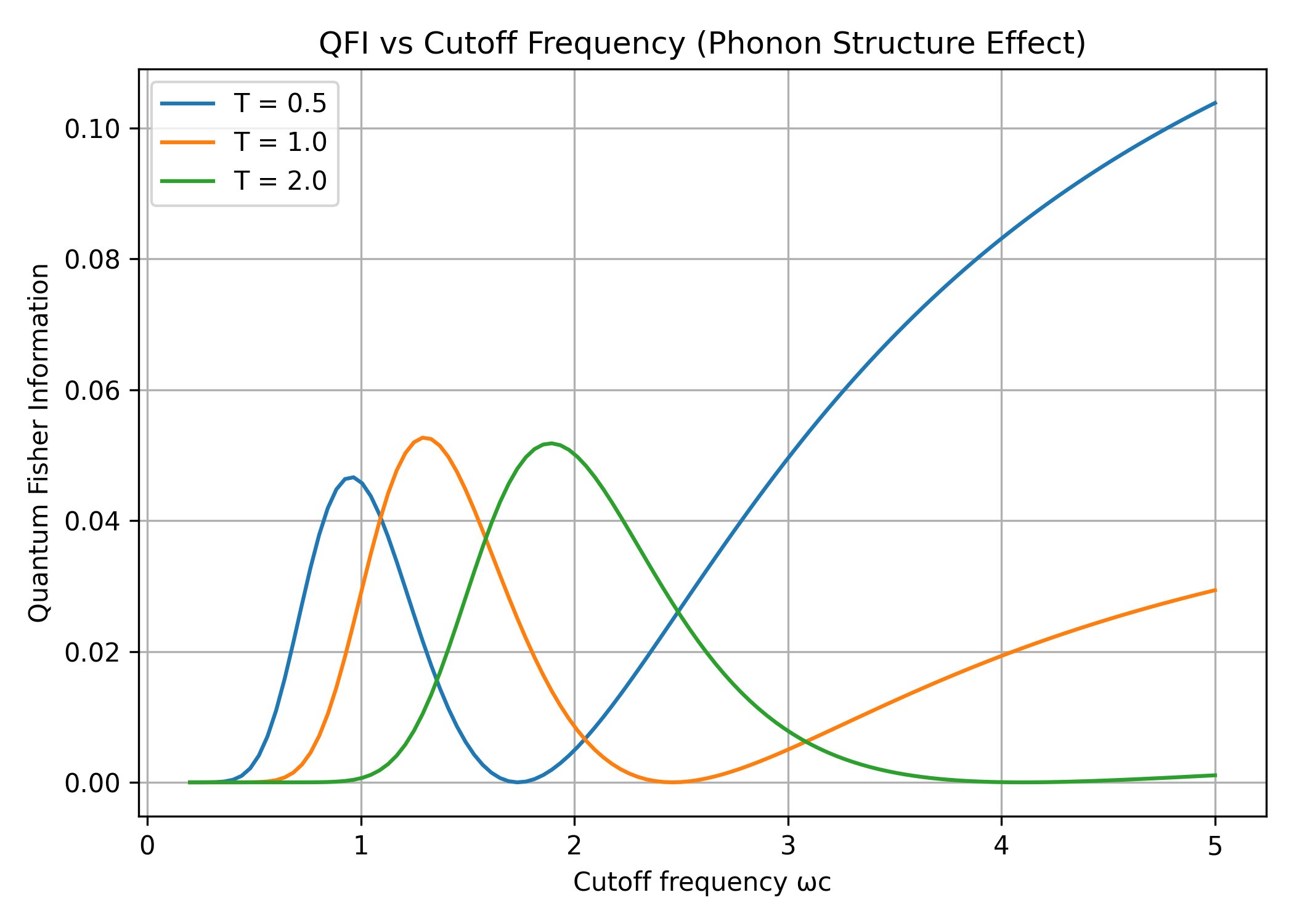}

\caption{Quantum Fisher Information (QFI) as a function of the environmental cutoff frequency $\omega_c$ for different temperatures. The QFI exhibits a non-monotonic dependence on $\omega_c$, with an optimal region that shifts with temperature, indicating the important role of the spectral structure of the environment in determining thermometric performance.}
\label{fig:qfi_wc}
\end{figure}

Figure~\ref{fig:qfi_wc} shows the dependence of the Quantum Fisher Information on the cutoff frequency $\omega_c$ for different temperatures. 

For small $\omega_c$, the spectral density suppresses high-frequency contributions, resulting in weak effective system–environment interaction and therefore low temperature sensitivity. As $\omega_c$ increases, the coupling to the phonon bath becomes more effective, which enhances the temperature dependence of the steady-state population and leads to an increase in the QFI. 

However, this increase is not monotonic. Instead, an optimal range of $\omega_c$ emerges where the QFI reaches its maximum. The position of this maximum shifts with temperature: at lower temperatures the optimal point appears at smaller $\omega_c$, while for higher temperatures it moves towards larger cutoff frequencies. 

This behavior indicates that the cutoff frequency is not merely a technical parameter, but plays an active role in controlling the metrological performance through the spectral properties of the environment.

\subsection{Limiting Case Analysis}

To further clarify the physical origin of the optimal sensitivity, we analyze several limiting regimes.

\paragraph{Weak-coupling limit ($\eta \to 0$).}

For small coupling strength, the phonon dressing factor approaches unity,
\begin{equation}
f(\eta,T) \approx 1,
\end{equation}
and the dissipation rate scales linearly,
\begin{equation}
\gamma \propto \eta.
\end{equation}

In this limit,
\begin{equation}
P_e \approx 
\frac{\Omega^2}
{2\Omega^2 + \omega_0^2},
\end{equation}
which is nearly independent of temperature. Consequently,
\begin{equation}
F_Q \to 0,
\end{equation}
indicating poor thermometric performance.

\paragraph{Strong-coupling limit ($\eta \to \infty$).}

For large coupling strength, the exponential dressing dominates,
\begin{equation}
f(\eta,T) \to 0.
\end{equation}

Both the effective frequency and dissipation rate are exponentially suppressed. The denominator of $P_e$ becomes dominated by $2\Omega^2$, leading to
\begin{equation}
P_e \to \frac{1}{2}.
\end{equation}

Since the population becomes temperature independent in this regime,
\begin{equation}
F_Q \to 0.
\end{equation}

\paragraph{Low-temperature limit ($T \to 0$).}

At low temperatures,
\begin{equation}
n(\omega_0,T) \sim e^{-\omega_0/T},
\end{equation}
and
\begin{equation}
\partial_T n(\omega_0,T)
\sim
\frac{\omega_0}{T^2} e^{-\omega_0/T}.
\end{equation}

 Despite the small occupation number at low temperatures, its derivative remains significant, enhancing temperature sensitivity. This explains the persistence of QFI enhancement at low temperatures in the intermediate-coupling regime.

These limiting cases confirm that the optimal sensitivity necessarily arises at an intermediate coupling strength, where neither linear dissipation nor exponential suppression fully dominates.
\subsection{Physical Implications and Future Prospects}
The non-monotonic behavior of the QFI observed in this study has important implications for the design of practical quantum sensors. Environmental coupling is often treated as a purely detrimental effect that leads to decoherence and loss of information. However, our results show that in the phonon-dressed regime, the environment can also play a constructive role by enhancing metrological sensitivity.

This suggests an alternative perspective: instead of trying to completely isolate the quantum probe, one can tune the system–environment interaction to reach an optimal intermediate coupling regime. In this sense, the environment becomes a controllable resource rather than a limitation.It is particularly instructive to contrast this with recent findings on the pulsed spectroscopy of emitters coupled to vibrational baths \cite{das2025}, where phonon-induced Franck-Condon suppression was found to monotonically reduce the precision for estimating internal parameters like the emitter's linewidth. Our work reveals a fundamentally different landscape for temperature estimation: the competition between dissipation and renormalization creates a window of enhanced sensitivity that is absent in spectroscopic tasks. This highlights that the impact of a structured phonon environment is not universally detrimental but depends critically on the specific parameter being estimated.

Our findings are consistent with recent studies indicating that nonequilibrium steady states and environmental engineering can improve thermometric performance compared to standard equilibrium approaches \cite{pekola2015, mitchison2020, neumann2013}.

From an experimental point of view, these results are particularly relevant for solid-state platforms such as nitrogen-vacancy (NV) centers in diamond and semiconductor quantum dots coupled to phononic environments. In such systems, it is possible to engineer the spectral properties of the environment, making it feasible to approach the optimal regimes identified in our analysis.

Future work could extend this framework to non-Ohmic environments or to many-body quantum probes, where collective effects may further enhance temperature sensitivity. Exploring these directions could provide additional insight into the role of structured environments in quantum metrology.
\newpage

\section{Conclusion and Discussion}

In this work, we have theoretically investigated the temperature sensitivity of a driven two-level quantum probe coupled to an Ohmic phonon environment. By utilizing a polaron transformation, we moved beyond the limits of weak-coupling approximations to account for the essential role of phonon-induced renormalization and dissipation.

Our main finding is the existence of a non-monotonic relationship between the Quantum Fisher Information (QFI) and the system-environment coupling strength $\eta$. We have demonstrated that the competition between environment-induced dissipation and the exponential suppression of the driving strength creates an optimal intermediate coupling regime where temperature sensitivity is maximized.

Crucially, our results offer a new perspective on the role of structured environments in quantum metrology. While recent studies on the spectroscopy of similar systems have shown that vibrational coupling monotonically degrades the precision for estimating internal parameters such as linewidth \cite{das2025}, we have shown that for thermometric tasks, the environment can be engineered as a constructive resource. This fundamental difference emphasizes that environmental effects must be evaluated specifically in the context of the target parameter.

It is worth noting that similar non-monotonic thermometric behavior has been reported in dephasing-based thermometry protocols \cite{Albarelli2023}
, where the sensitivity is governed by the interplay between information extraction and measurement-induced disturbance (invasiveness). In such approaches, the temperature typically refers to that of a finite sample, and the thermometric performance is closely related to heat exchange and back-action effects. In contrast, the present work adopts an open quantum systems framework to estimate the temperature of a surrounding phonon environment, treated as an extended thermal reservoir that effectively represents the external (ambient) temperature of the system. Consequently, the optimal sensitivity identified here originates from the competition between phonon-induced renormalization and dissipation, rather than from information--disturbance trade-offs.

Our findings provide a practical guide for optimizing solid-state quantum thermometers, such as NV centers and quantum dots, where the spectral properties of the phonon bath can be tuned. Future investigations could explore the impact of non-Ohmic spectral densities or multi-qubit probes to further enhance the limits of quantum thermal sensing in complex environments.

\section*{Acknowledgements}
The author expresses her gratitude to  Animesh Datta for providing the research environment and for the helpful discussions. This work was supported by The Scientific and Technological Research Council of Türkiye (TÜBİTAK) under the 2219-International Postdoctoral Research Fellowship Program. The hospitality and support provided by the Department of Physics at the University of Warwick are also gratefully acknowledged.

\newpage
\section*{Appendix A: Polaron Transformation Details}

To derive the effective Hamiltonian and the renormalized decay rates, we apply the unitary transformation:
\begin{equation}
H' = U H U^{\dagger}
\end{equation}
The displacement operator effectively dresses the qubit with a cloud of phonons. 
Applying this transformation explicitly, the effective system-bath Hamiltonian becomes:
\begin{equation}
H' = \frac{\omega_0}{2}\sigma_z + \Omega \sigma_x \cosh(B) + i\Omega \sigma_y \sinh(B) + \sum_k \omega_k b_k^\dagger b_k,
\end{equation}
where $B = \sum_k (\lambda_k b_k^\dagger - \lambda_k^* b_k)$ is the bath displacement operator. Tracing over the thermal bath yields the renormalized system parameters $\omega_{eff}$ and the modified drive.

The second-order expansion of the Master equation in the polaron frame leads to the non-Markovian rates. Specifically, the time-dependent correlation functions of the bath are integrated:
\begin{equation}
C(t) = \int_0^{\infty} d\omega J(\omega) [n(\omega,T)e^{i\omega t} + (n(\omega,T)+1)e^{-i\omega t}]
\end{equation}

\vspace{1cm}

\section*{Appendix B: QFI for a Diagonal Steady State}

For a generic state
\begin{equation}
\rho = \sum_i P_i |i\rangle \langle i|
\end{equation}
the Symmetric Logarithmic Derivative (SLD) $L_T$ is defined by
\begin{equation}
\partial_T \rho = \frac{1}{2}(\rho L_T + L_T \rho)
\end{equation}
In the energy basis where $\rho$ is diagonal, the elements of $L_T$ are given by
\begin{equation}
(L_T)_{ii} = \frac{\partial_T P_i}{P_i}
\end{equation}
The QFI is then calculated as
\begin{equation}
F_Q = \text{Tr}(\rho L_T^2) = \sum_i \frac{(\partial_T P_i)^2}{P_i}
\end{equation}
This confirms that the population-based Fisher Information saturates the quantum bound for our specific steady-state configuration.

\end{document}